# Optoelectronic spin memories of electrons in semiconductors

M. Idrish Miah[1,2]

**Abstract** We optically generate electron spins in semiconductors and apply an external magnetic field perpendicularly to them. Time-resolved photoluminescence measurements, pumped with a circularly polarized light, are performed to study the spin polarization and spin memory times in the semiconducting host. The measured spin polarization is found to be an exponential decay with the time delay of the probe. It is also found that the spin memory times, extracted from the polarization decays, enhance with the strength of the external magnetic field. However, at higher fields, the memory times get saturated to sub-$\mu$s because of the coupling for interacting electrons with the local nuclear field.

## Introduction

Spintronics is a relatively new and emerging field in solid-state physics where, rather than the charge of the electron, its spin plays the dominant role. The study of the spin property of electrons in solid-state systems and its exploitation for future technological applications is the main task of spintronics (Prinz 1998; Dyakonov and Khaetskii 2008; Ziese and Thornton 2001; Awschalom et al. 2002). However, for the successful incorporation of spins into the currently existing semiconductor technology, one has to resolve technical issues such as efficient generation/injection of spins and their optimal control or spin dynamics (Awschalom et al. 2002).

Generation/injection of spins (or spin polarization) usually means creating a non-equilibrium spin population. This has been achieved either by optical method (using a circularly polarized light excitation) or by an electrical means by magnetic semiconductors, or ferromagnetic contacts (Prinz 1998; Awschalom et al. 2002). Although electrical spin injection is desirable, this technique is found not to be efficient and has resulted in low spin injection effects due to the conductivity mismatch. However, the spin generation by the optical methods has been successful (Prinz 1998) and the high spin polarization of conductor band electrons in semiconductor heterostructures has been obtained (Endo et al. 2000).

Despite substantial progress in optical spin generation, a further hurdle still remains in the spin transport which is the lack of a proper understanding of spin dynamics and control in semiconductor-based heterostructures (Gotoh et al. 1998; Cortez et al. 2002; Wang et al. 2007). Spin dynamics in GaAs-based semiconductors and their low-dimensional systems has been studied by time-resolved polarization of photoluminescence (PL) (Seymour and Alfano 1980; Wagner et al. 1993; Endo et al. 2000; Gotoh et al. 1998; Cortez et al. 2002). The PL spin polarization was used to directly measure the spin-flip times which correspond to the spin memory times ($\tau_s$) in the samples. The measured values range from 100 ps to 20 ns in p-doped GaAs and related materials (Seymour and Alfano 1980; Wagner et al. 1993; Endo et al. 2000), and 1–15 ns in InGaAs quantum wells and n-type InAs/GaAs quantum

✉ M. Idrish Miah
m.miah@griffith.edu.au

1 Department of Physics, University of Chittagong, Chittagong 4331, Bangladesh

2 Queensland Micro- and Nanotechnology Centre, Griffith University, Nathan, Brisbane, QLD 4111, Australia

dots (Gotoh et al. 1998; Cortez et al. 2002). Here, in the present investigation, electron spins in lightly *n*-doped GaAs are generated optically in the presence of an external magnetic field. The detection of them is performed using a time-resolved pump-probe excitonic PL spin polarization measurement.

## Experimental details

Samples were GaAs layers in an AlGaAs heterostructure, grown by molecular beam epitaxy, and were lightly *n*-doped with a doping density of $3 \times 10^{15}$ cm$^{-3}$. A circularly polarized light from a modulated Ti–sapphire laser, operated at 30 mW and 810 nm, was used to generate spin-polarized electrons in the samples. The laser intensity was modulated on/off with an acousto-optic modulator (AOM) to obtain light pulses as short as 15 ns. The AOM was controlled by an Interface Technology via digital word generator (DWG), which delivered voltage pulses to the AOM's pulse shaping input. The DWG was controlled by a pulse controller, and was triggered by a 20 kHz photo-elastic modulator (PEM) in the PL detection path. The PEM operated as a sinusoidally oscillating quarter-wave plate, which was combined with a linear polarizer to make a circularly polarization analyzer. The PEM additionally triggered the two channels of the counter so that the two ($\sigma^+$, right and $\sigma^-$, left) circular polarizations could be separately recorded. For the application of the magnetic field (B) in the Voigt configuration, an Oxford superconducting magnet was used. Experiments were performed at low temperature. The sample temperature was measured, and it was 4.2 K. A SPEX 1680 double grating spectrometer, with a photomultiplier tube (PMT) and photon counter/quantum sensor (LI-190 SA, LI-COR Inc.), was used to collect and measure the PL. Important characteristics of the present experiment include using excitonic PL polarization to monitor the spin polarization of doped electrons, initializing the electronic polarization with a lengthy pump light pulse, and reading out the electronic polarization with a shorter probe light pulse (probe), where the spin polarization relaxes during the dark period between pump and probe pulses. The degree of PL circular polarization (*P*) was calculated by defining it as the ratio of the difference of the PL emission intensities of the right and left circularly polarized PL to their sum, i.e.,

$$P = \frac{I_{PL+} - I_{PL-}}{I_{PL+} + I_{PL-}},$$

where $I_{PL+}$ ($I_{PL-}$) is the PL emission intensity of the right (left) circularly polarized PL. The PL emission intensities were measured in the left and right circular polarizations under the right circularly polarized excitation.

## Results and discussion

The pump pulse generated spin-polarized electrons, while a probe pulse read out their polarization. Figure 1 shows typical PL emission intensities measured using pulses with the same and opposite circular polarizations. There is a difference between the different spin polarization conditions, which is caused by spin-dependent phase-space filling (Wang et al. 2007). From the measured data, the PL integrated intensity was used for the calculations of the PL polarization.

In order to explore conditions for pump and probe pulses, we also studied a dependence of the circular polarization on the length of a single pulse for a constant pump power density. It was found that the polarization increases with the increase in the pulse length, but for the higher pulse lengths, it gets saturated. The free exciton line is polarized to a degree that depends on the pulse length. At a small pulse length, the polarization is small and depends on the spin relaxation, electron density, and generation rate, whereas at larger pulse lengths, the polarization saturates while the power density is held constant. The number of photo-generated electrons must be comparable to the number of doped electrons in order for an appreciable polarization to be set by the light pulse. The results suggest a measurement using a probe pulse smaller than 40 ns and a pump larger than 220 ns. The present pump-probe measurements were performed with a delay ($\Delta\tau$) using 250 ns pump pulses and 25 ns probe pulses. In the measurement conditions, the quantum sensor ensured that PL was only collected from the probe pulse.

Figure 2 shows the spin polarization (mean data point) as a function of $\Delta\tau$ for different strengths of the magnetic field. As obvious, the PL polarization decays with $\Delta\tau$. The observed PL polarization decay corresponds to electron spin relaxation in the sample for the transverse magnetic field. For a long delay, there is still a small polarization value. This is because the probe pulse cannot measure the system without affecting it, as it is circularly polarized (Miah 2011a). However, the probe must be weak enough that one can observe the existing polarization, but strong enough so that it produces enough photoluminescence yield to detect rigorously. For this reason, for example, for less-doped samples, weaker probe beams are required, but there might be a difficulty for detecting the signals. For shorter delays, a larger polarization was seen, demonstrating that polarization persisted into the dark period between pulses. The polarization decayed exponentially between the short- and long-time limits in accordance with the decay law. The experimental data are well fitted by the exponential decay. Exponential fits to the data give the spin-flip times, corresponding to the spin memory times for the respective fields.

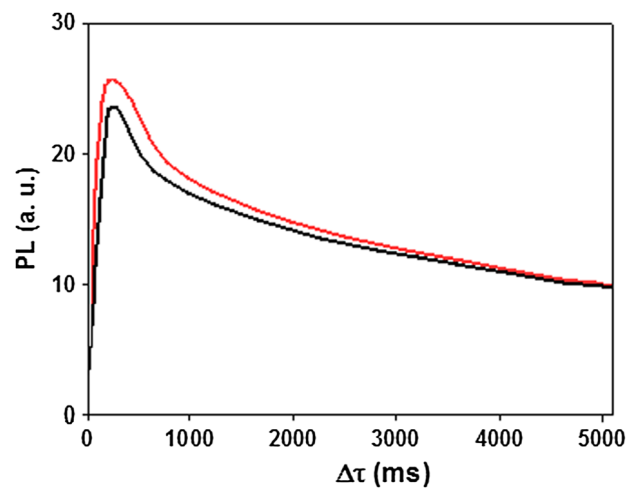

**Fig. 1** PL intensities measured using optical pulses with the same (*upper*) and opposite (*lower*) circular polarizations

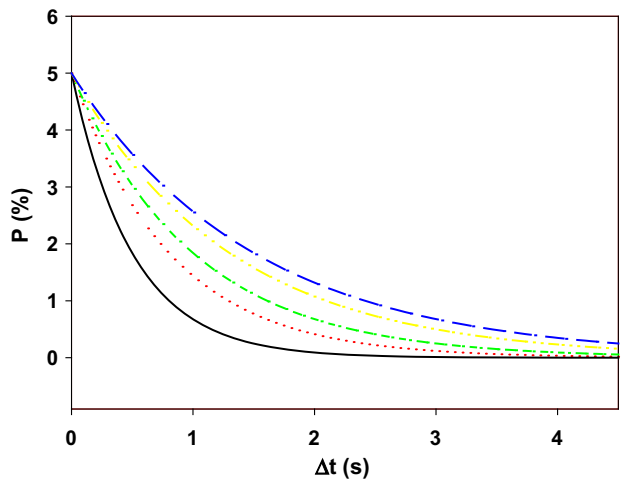

Fig. 2 Spin polarization (mean data) as a function of the pump-probe delay for different magnetic field strengths: 0.5 T (*solid line*), 0.9 T (*dotted line*), 1.6 T (*short dashed line*), 3 T (*dot-dashed line*), and 5 T (*long dashed line*)

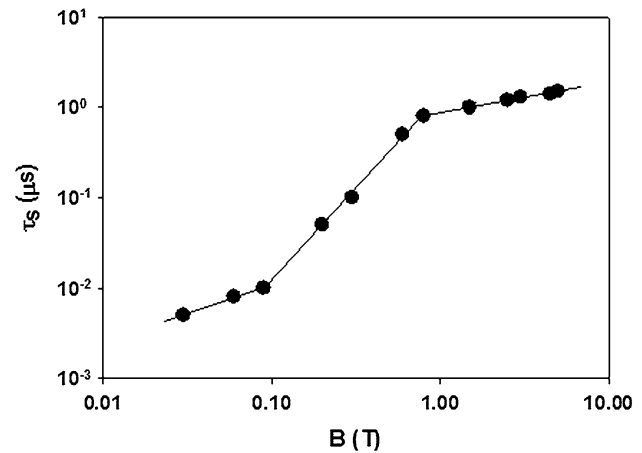

Fig. 3 Magnetic field dependence of the spin memory time

The dependence of the spin memory times on the magnetic field is shown in Fig. 3. As can be seen, the spin memory time increases with the increase in the strength of the magnetic field up to about 8 T. However, at higher magnetic fields, it got saturation. In the intermediate field strengths, there is also another feature where a rapid increase of the memory time is seen. This regime corresponds to interacting electrons at slightly higher fields, because there is a natural distribution of donor separations, which can lead to more- and less-localized electrons (Pikus and Titkov 1984; Kityk 1991). This means that the high-field regime corresponds to the localized electrons. For localized electrons, the main relaxation mechanism under this condition is hyperfine coupling to the nuclei. The hyperfine interaction produces a fluctuation magnetic field, also known as the effective magnetic field, in which an electron precesses (Prinz 1998). As an external magnetic field is applied, the nuclear contribution to spin relaxation will be reduced when the external field exceeds the nuclear fluctuation field (Lampel and Weisbuch 1975; Ivchenko and Kiselev 1992).

However, the lifetime at zero fields should be equal to the inhomogeneous spin dephasing time, since there is no energy splitting between the two spin states. The spin relaxation decreases with the increase in strength of the magnetic field for the lower-field region. For the intermediate region, the motional averaging occurs and in this regime, spin memory time will increase with the external magnetic field with the spin relaxation (Pikus and Titkov 1984). This field dependence arises from motional averaging of the hyperfine effects for interacting electrons. However, the application of a magnetic field in the transverse configuration tends to localize electrons due to cyclotron motion (Dzhioev et al. 1997). As the external magnetic field is further increased, the motional averaging is no longer a constant and increases as the electrons become localized due to the field (Salimullah et al. 2003). As a result, the precession frequency increases with the field and becomes comparable to the inverse of the correlation time corresponding to the motional averaging at the fields of some T (Miah 2011b).

## Conclusions

Spin memories of electrons in optoelectronic semiconductor devices were investigated. The dynamics of spins generated by an optical technique was studied in the presence of an external magnetic field applied perpendicularly to the spins. A time-resolved pump-probe photoluminescence measurement was performed to study the spin polarization and spin memory times in the host. The measured spin polarization was found to be an exponential decay with the time delay. It was also found that the spin memory times enhance with the strength of the externally applied transverse magnetic field. However, because of the hyperfine coupling for interacting electrons with the local nuclear field (an interaction between the electron spin and the nuclear spin), the spin memory times were found to get saturated at higher fields. The findings resulting from this investigation might have potential applications in semiconductor-based opto-spintronics devices.